\documentclass[a4paper,11pt]{article}
\usepackage{pos}

\title{Constraints on jet quenching from a multi-stage energy-loss approach}

\manuallySeparateAuthors 
\author*{Chanwook Park}
\author{ for the JETSCAPE collaboration}

\affiliation{Department of physics, McGill University,\\
  3600 University Street, Montr\'{e}al, QC, H3A 2T8, Canada}

\emailAdd{chanwook@physics.mcgill.ca}

\abstract{
We present a multi-stage model for jet evolution through a quark-gluon plasma within the JETSCAPE framework.
The multi-stage approach in JETSCAPE provides a unified description of distinct phases in jet shower contingent on the virtuality.
We demonstrate a simultaneous description of leading hadron and integrated jet observables as well as jet $v_n$ using tuned parameters.
Medium response to the jet quenching is implemented based on a weakly-coupled recoil prescription.
We also explore the cone-size dependence of jet energy loss inside the plasma.
}

\FullConference{%
  HardProbes2020\\
  1-6 June 2020\\
  Austin, Texas}

\begin{document}
\maketitle

\section{Introduction}

Jet evolution through the QGP is characterized by several distinct phases depending on jet virtualities, and different energy loss mechanisms are essential to describe each stage.
A multi-stage approach within the JETSCAPE framework provides a unified description of the jet shower, including a high-virtuality gluon-splitting phase and a low-virtuality scattering-dominated phase.
In these proceedings, we report a comprehensive study of multi-stage jet evolution by performing a model-to-data comparison to constrain the jet quenching parameter in heavy-ion collisions.

\section{Unified approach in JETSCAPE}

Throughout this study, a dynamically evolving QGP created in Pb-Pb collisions at $\sqrt{s_{NN}} = 5.02$ TeV is simulated using ($2+1$)-D VISHNU~\cite{Shen:2014vra} with fluctuating T$\raisebox{-0.4ex}{R}$ENTo~\cite{Moreland:2014oya} initial conditions, followed by free-streaming and dissipative fluid dynamics.
Hard partons produced by PYTHIA~\cite{Sjostrand:2014zea} with initial state radiation (ISR) and multi-parton interaction (MPI) are initialized in the transverse plane by T$\raisebox{-0.4ex}{R}$ENTo profiles for initial binary collisions.
These partons then evolve through the hydrodynamic medium.
The multi-stage energy loss formalism consists of MATTER~\cite{Majumder:2013re,Abir:2015hta} for the high-virtuality stage and LBT~\cite{Wang:2013cia,He:2015pra} for the low-virtuality stage.
The phase spaces for the two energy loss models are separated by a switching virtuality $Q_0$.
The simulation of p+p collisions is performed by MATTER vacuum showers using the JETSCAPE PP19 tune~\cite{Kumar:2019bvr}.

MATTER is a Monte-Carlo event generator for partons with virtuality $Q > Q_0$.
Parton splittings are described by a generalized Sudakov form factor, which includes vacuum and medium-modified parton splitting functions.
The in-medium contribution, which induces transverse momentum broadening of jets, $\hat{q}$, in a QGP, is estimated based on the Higher-Twist energy loss model~\cite{Wang:2001ifa,Majumder:2009ge,Qin:2009gw}.
We have used a hard thermal loop technique~\cite{CaronHuot:2010bp} to formulate $\hat{q}$.

The time-ordered in-medium shower in LBT for low virtuality partons relies on solving a linearized Boltzmann equation with in-medium kernels.
The model contains leading order $2 \rightarrow 2$ elastic and $2 \rightarrow 2 + n$ inelastic scatterings, where $n$ indicates multiple gluon radiation.
The Higher-Twist formalism evaluates the average number of emitted gluons from a hard parton, which follows the Poisson distribution.

The switching virtuality $Q_0$ is set to $1$, $2$, and $3$ GeV, and a value of $\alpha_s = 0.25$ is used for the strong coupling to determine the quenching parameter $\hat{q}$.
Our previous analysis of the single hadron and jet nuclear modification factor $R_{AA}$ constrained these model parameters~\cite{Kumar:2020vkx}. 
Both the MATTER and the LBT in-medium showers implemented recoil partons based on a weakly-coupled picture to reproduce the medium response to jet quenching.
The energies and momenta originating from incoming thermal partons during jet-medium scattering (holes) are subtracted from the jet signals in the final state.

\section{Results}

\begin{figure}[t]
	\centering
	\includegraphics[width=0.75\textwidth]{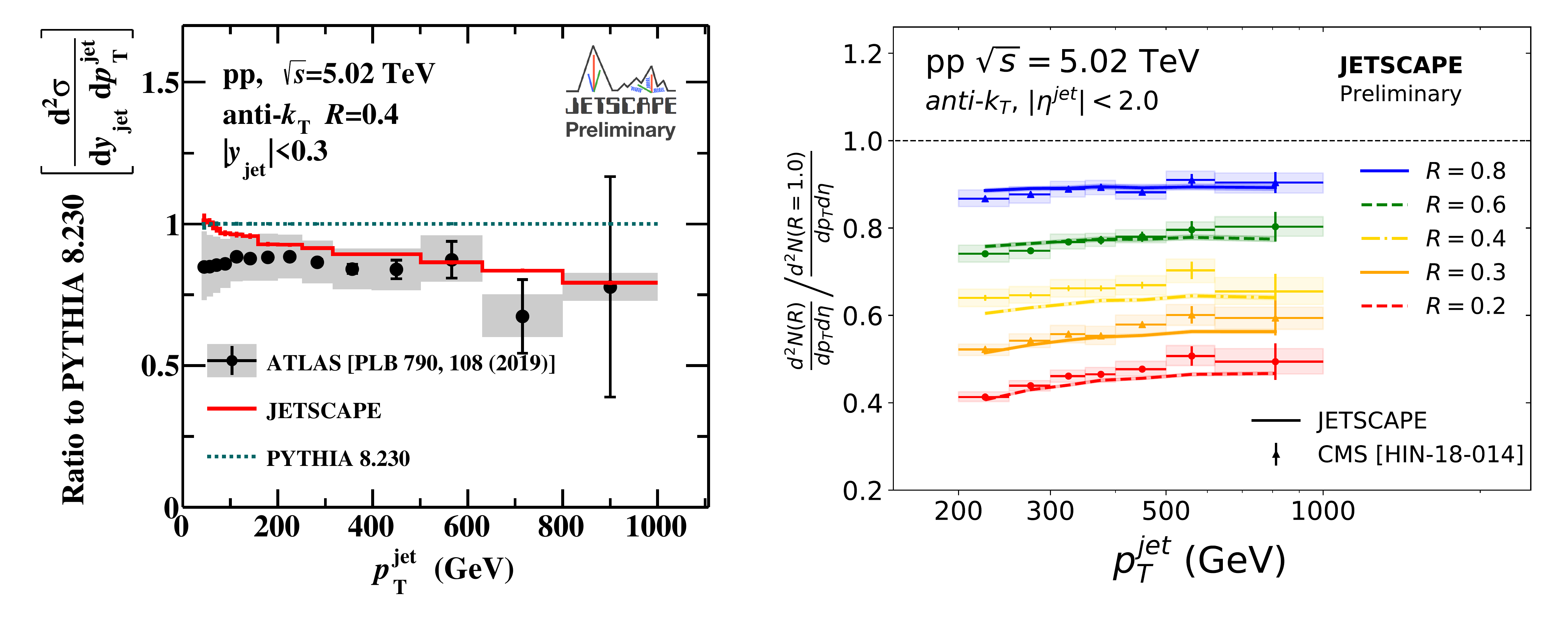}
	\caption{Comparison between the results obtained from the JETSCAPE PP19 tune at $\sqrt{s_{NN}} = 5$ TeV and measurements. (Left) Inclusive jet cross-section with $|y_{jet}| < 0.3$~\cite{Aaboud:2018twu}, normalized by the PYTHIA predictions. (Right) Ratio of the jet spectra for $R = 0.2$ to $0.8$ with respect to $R = 1.0$~\cite{CMS:2019btm}.
	}
	\label{fig:PP19tune}
\end{figure}

\begin{figure}[t]
	\centering
	\includegraphics[width=0.75\textwidth]{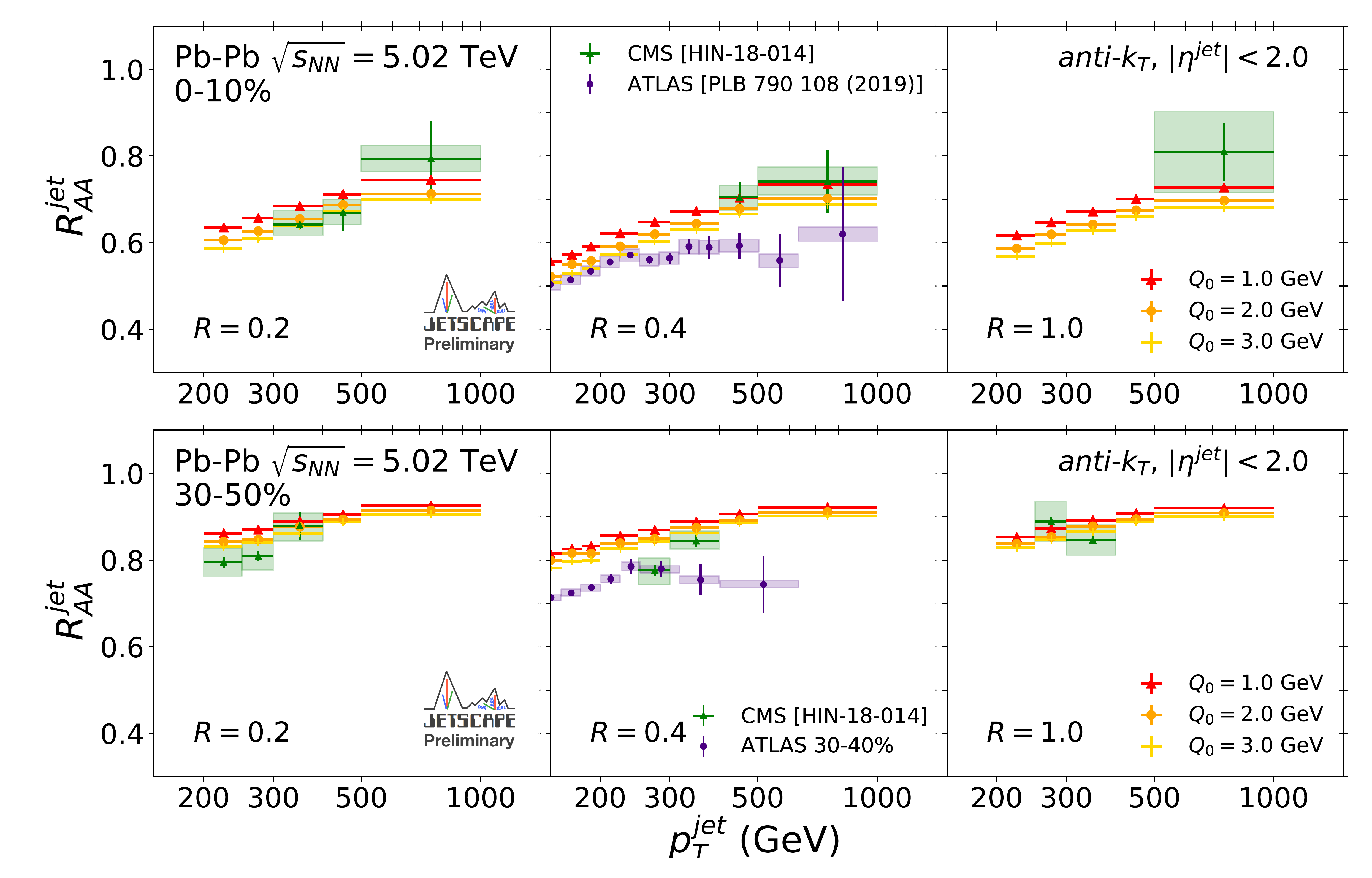}
	\caption{Inclusive jet $R_{AA}$ in central (top panel) and peripheral (bottom panel) Pb-Pb collisions at $\sqrt{s_{NN}} = 5.02$ TeV with various $R$ and $Q_0$ values~\cite{CMS:2019btm}.
	}
	\label{fig:jet_raa}
\end{figure}

\begin{figure}[t]
	\centering
	\includegraphics[width=0.9\textwidth]{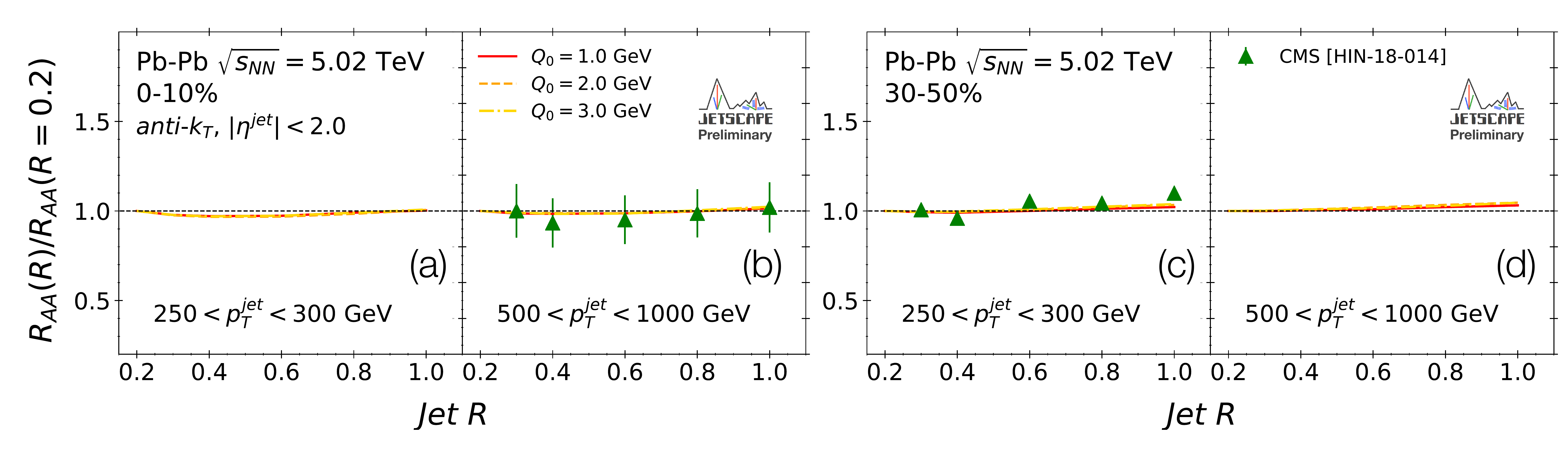}
	\caption{Ratio of jet $R_{AA}$ as a function of $R$ with respect to $R = 0.2$ in central (a-b) and peripheral (c-d) Pb-Pb collisions with two jet $p_T$ intervals.
    The data is calculated from the jet $R_{AA}$ results shown in Fig.~7 in~\cite{CMS:2019btm}.
	}
	\label{fig:ratio_jet_raa}
\end{figure}

\begin{figure}[t]
	\centering
	\includegraphics[width=0.75\textwidth]{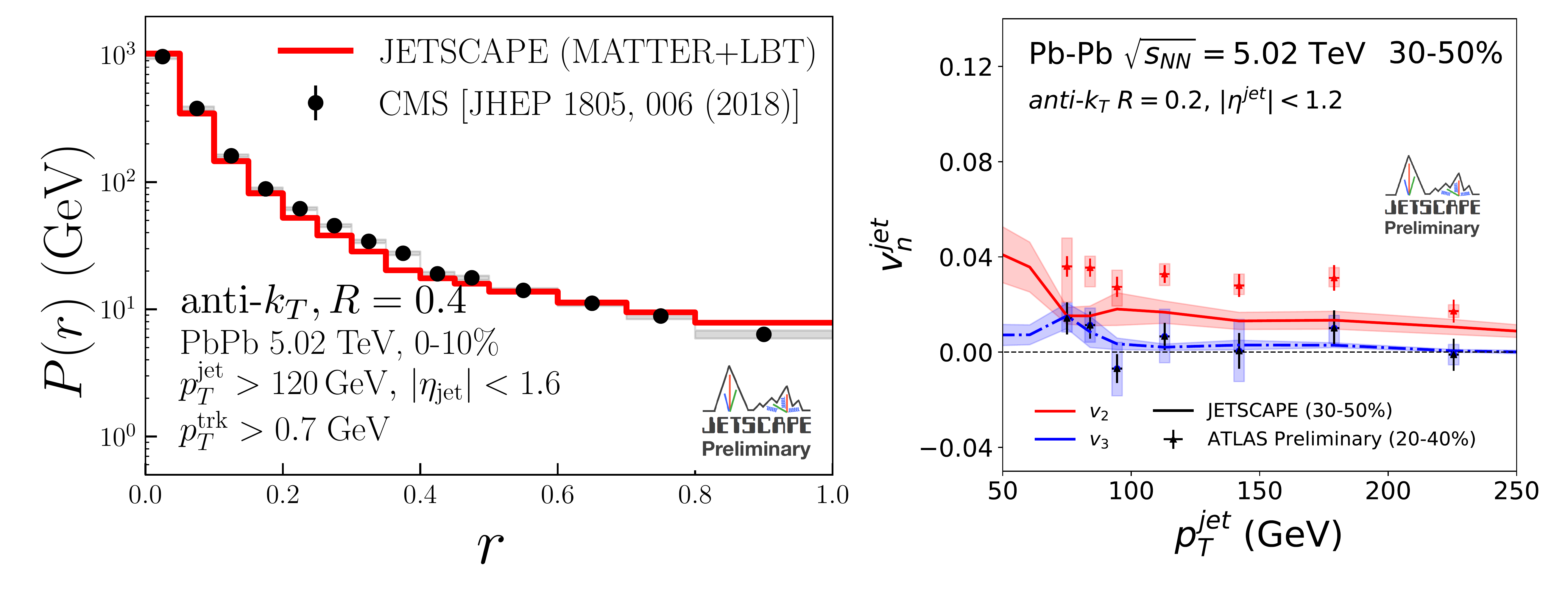}
	\caption{(Left) Jet shape function for $R = 0.4$ jets in central Pb-Pb collisions~\cite{Sirunyan:2018jqr}. (Right) Anisotropic flow coefficients $v_2$ and $v_3$ for jets at peripheral Pb-Pb collisions~\cite{ATLAS-CONF-2020-019}.
	}
	\label{fig:jet_shape_and_vn}
\end{figure}

The left panel of Fig.~\ref{fig:PP19tune} shows the jet cross section in p+p collisions at $\sqrt{s_{NN}} = 5.02$ TeV with two rapidity cuts, normalized by the PYTHIA predictions.
The $p_T$ dependence of the jet cross-section is consistent with data for jet $p_T > 200$ GeV at mid-rapidity. 
The ratio of the jet cross-section with various $R$ with respect to $R = 1.0$ is displayed in the right panel in Fig.~\ref{fig:PP19tune}.
The angular dependence of the jet $R_{AA}$ is well reproduced by the MATTER vacuum shower in JETSCAPE with the PP19 tune.

We present the jet $R_{AA}$ with various $R$ and switching virtualities $Q_0$ in central and peripheral Pb+Pb collisions in Fig.~\ref{fig:jet_raa}.
We consistently observe stronger jet quenching with larger values of $Q_0$.
The parton shower in the high-virtuality phase (MATTER) is dominated by virtuality splitting, but the low-virtuality phase (LBT) is largely affected by scatterings, which induce jet $p_T$ broadening.
This accounts for the jet $R_{AA}$ being more suppressed when the LBT phase starts at higher virtuality $Q_0$.
The jet $R_{AA}$ independent to $R$ leads to the $R_{AA}$ ratio with respect to $R = 1.0$ consistent with unity unity as shown in Fig.~\ref{fig:ratio_jet_raa}.
This monotonic behavior is independent of centrality and jet $p_T$, implying that the jet energy contained within $R < 0.2$ generally dominates the jet $R_{AA}$ value.
The steeply falling jet shape function shown in the left panel of Fig.~\ref{fig:jet_shape_and_vn} supports this interpretation.
However, a rigorous investigation of recoils would be necessary as their influence on jet shape is expected to be significant at larger $R$.

The right panel in Fig.~\ref{fig:jet_shape_and_vn} shows the jet $v_2$ and $v_3$ in peripheral Pb-Pb collisions.
The observed non-zero $v_2$ for high-energy jets originates from the path-length dependent jet quenching in an almond-shaped QGP.
The vanishing jet $v_3$ within the statistical uncertainties is consistent with the data.

\section{Conclusion}

We have studied jet modification using a unified approach within the JETSCAPE framework.
The results for the jet cross-section in pp collisions using the JETSCAPE PP19 tune show good agreement with data.
The multi-stage model with a combination of MATTER and LBT provides a simultaneous description of the integrated and differential jet observables.
Our future work will investigate recoils for the detailed jet quenching mechanism at large $R$.


\end{document}